\documentclass[11pt]{article} 
\usepackage{geometry,amsmath,amssymb,bm,graphicx}
\usepackage[sort&compress,numbers]{natbib} 
\usepackage{amsmath,amssymb,bm}
\usepackage{url}
\usepackage{mystyle}

\newcommand{\tm}{\mathsf{tm}}
\newcommand{\te}{\mathsf{te}}
\newcommand{\gr}{\mathsf{gr}}

\newcommand{\ii}{\mathrm{i}}
\newcommand{\id}{\mathrm{id}}
\newcommand{\dd}{\mathrm{d}}
\newcommand{\etaB}{\bm{\eta}}
\newcommand{\RC}{\bm{\mathcal{R}}}
\newcommand{\rB}{\bm{r}}
\newcommand{\RB}{\bm{R}}
\newcommand{\uB}{\bm{u}}
\newcommand{\kB}{\bm{k}}
\newcommand{\tB}{\bm{t}}
\newcommand{\TBB}{\bm{T}}
\newcommand{\IB}{\mathbf{I}}
\newcommand{\GB}{\mathbf{G}}
\newcommand{\TB}{\mathbf{T}}
\newcommand{\sigmaB}{\bm{\sigma}}
\newcommand{\vB}{\bm{v}}
\newcommand{\nB}{\bm{n}}
\newcommand{\kT}{\tilde{k}}
\newcommand{\PhiT}{\widetilde{\Phi}}
\DeclareMathOperator{\tr}{tr}
\DeclareMathOperator{\sign}{sign}
\DeclareMathOperator{\arctanh}{arctanh}

\newcommand{\bs}{\begin{subequations}}
\newcommand{\es}{\end{subequations}}
\newcommand{\Vl}[1]{\stackrel{_\leftarrow}{#1}}
\newcommand{\Vr}[1]{\stackrel{_\rightarrow}{#1}}
\newcommand{\Eb}{\mathbf{E}}

\usepackage{xcolor} \definecolor{darkgreen}{rgb}{0,.5,0}
\usepackage[colorlinks,filecolor=blue,citecolor=darkgreen,unicode]{hyperref}

\title{The normal Casimir force for lateral moving planes with isotropic conductivities}

\author{N. Emelianova and N. Khusnutdinov\footnote{email: nail.khusnutdinov@gmail.com}\\[1em]
	\small CMCC, Universidade Federal do ABC, Avenida dos Estados 5001, CEP 09210-580, SP, Brazil}
\date{\small January 2, 2024}
\begin{document}
	\maketitle

\begin{abstract}
	We consider the two planes at zero temperature with isotropic conductivity that are in relative lateral motion with velocity $v$ and inter-plane  distance $a$. Two models of conductivity are taken into account -- the constant and frequency-dependent Drude models. The normal (perpendicular to planes) Casimir force is analysed in detail for two systems -- i) two planes with identical conductivity and ii) one of the planes is a perfect metal. The velocity correction to the Casimir energy $\Delta_v\mathcal{E} \sim v^2$ for small velocity for all considered cases. In the case of the constant conductivity $\eta$, the energy correction is $ \Delta_v\mathcal{E} \sim \frac{\eta}{a^3} \left(\frac{v}{\eta}\right)^2$for $v\ll \eta \ll 1$.
\end{abstract}

\section{Introduction}

The Casimir effect was initially considered for perfect conductive plates, and nowadays, it is extended to many non-ideal and new materials \cite{Bordag:2009:Ace,Woods:2016:MpCvWi}. Casimir noted in Ref.\,\cite{Casimir:1998:SrhscCe} that Bohr suggested considering the zero-point energy as an origin of this effect and simplifying the derivation of the force. In the case of perfect conductive planes, the Casimir effect relies solely on fundamental constants and interplane distance. However, for actual materials, the Casimir effect depends on various factors, such as the shape and structure of the material, conductivity, chemical potential, temperature, and presence of impurities, etc. \cite{Bordag:2009:Ace,Woods:2016:MpCvWi}. 

The Casimir force between bodies is further influenced by their relative motions (see the recent review on the dynamic Casimir effect \cite{Dodonov:2020:FYDCE} and Refs.\,\cite{Bordag:2009:Ace,Bordag:2001:NdCea,Khusnutdinov:2019:cei2dmss}).  The relative motions are lateral (parallel to the planes), perpendicular to planes, or, in general, their combinations.  The Casimir effect for perpendicularly and uniformly moving slabs has been considered firstly in Refs.\,\cite{Bordag:1984:Ccesfsnbc,Bordag:1986:CEUMM} for electromagnetic and massless scalar fields. It is a direct consequence of the Quantum Field Theory with the moving boundaries \cite{Fulling:1976:Rmmtdsca}. In the non-relativistic scenario, the velocity correction to the Casimir pressure is quadratic $\sim v^2$ ($\sim v^2/c^2$ in dimensional units) for both fields, but with opposite signs. For the scalar field, the relative velocity correction for Casimir pressure $\delta\mathcal{P} = (\mathcal{P} - \mathcal{P}_{v=0})/\mathcal{P} = \frac{8}{3}v^2$, while for electromagnetic field $\delta\mathcal{P} = -\left(\frac{10}{\pi^2} - \frac{2}{3}\right)v^2$. It is positive for the massless scalar field and it is negative for the electromagnetic case. 

The lateral relative motion of different planes gives rise to two distinct Casimir pressures in perpendicular directions. One of these pressures acts normal to planes, similar to perpendicular motion, while the other acts along planes, known as quantum, non-contact, or Casimir friction.  The study of the normal force has been carried out in previous works \cite{Maslovski:2011:Crmm,Philbin:2009:Nqfump} for layers in stratified dielectric media with magneto-electric and non-reciprocal constant coupling and plates with general electric permittivity and magnetic permeability, respectively. For a three-layer system \cite{Maslovski:2011:Crmm}, the force can be either attractive or repulsive depending on the velocity directions of the extreme layers. In the non-relativistic case, the velocity correction to the force becomes repulsive when the extreme layers have the same velocity directions with respect to the middle layer, and attractive for opposite velocity directions.The velocity correction to the Casimir energy follows a similar order of magnitude, approximately proportional to $v^2$.  For relativistic velocities, both attractive and repulsive effects can occur.  A general expression for the normal force between two plates was obtained in Ref.\,\cite{Philbin:2009:Nqfump} using the Fresnel reflection coefficients. It was shown that the same quadratic correction $\sim v^2$ applies to the usual Casimir force.

Quantum friction is a more challenging topic for analysis and is currently a subject of debate \cite{Polevoi:1990:Tmfcmbfef,Dedkov:2018:FrhestppmsLPRtr,Mkrtchian:1995:Immbvev,Pendry:1997:Svqf,Volokitin:2007:Nrhtnf,Philbin:2009:Nqfump,Pendry:2010:Qfff,Maghrebi:2013:SadCe,Maghrebi:2013:QCrnf,Farias:2017:Qfgs,Farias:2018:Qfttm,Brevik:2022:Feoe,Guo:2023:Qfppcp,Antezza:2023:Cfmg}, with some even negating its existence \cite{Philbin:2009:Nqfump}. Two dielectric planes at different temperatures with lateral relative motion have been considered in Refs.\,\cite{Polevoi:1990:Tmfcmbfef,Dedkov:2018:FrhestppmsLPRtr}. The quantum friction force was calculated within the framework of the Rytov fluctuation theory   \cite{Levin:1967:TETFE}. It was demonstrated that the force is proportional to the first power of the velocity $v$, but can have different signs, resulting in either deceleration or acceleration of the planes.  Mkrtchian investigated two conductive planes with relative lateral motion and calculated the force and viscosity of vacuum for different plane impedance models \cite{Mkrtchian:1995:Immbvev}. The dependence of the force on the inter-plane distance heavily relies on the chosen model, but in any case, the velocity correction to the usual Casimir force $\sim v$  emerges as the typical dependence for Casimir friction. The quantum friction for planes with temporally dispersive conductivity was analyzed in Ref.\,\cite{Pendry:1997:Svqf}, revealing a $\sim v^3$ dependence at small velocities and a $\sim v^{-1} \ln v$ dependence at high velocities. Volokitin and Persson \cite{Volokitin:2007:Nrhtnf} developed a general theory for quantum friction within the framework of fluctuation electrodynamics. While Ref.\,\cite{Philbin:2009:Nqfump} claims the absence of quantum friction as a whole, the existence of a normal force remains without doubt (see Ref.\,\cite{Pendry:2010:Qfff}). 

The study of quantum friction using scattering theory was carried out in Refs.\,\cite{Maghrebi:2013:SadCe,Maghrebi:2013:QCrnf}, demonstrating the existence of a quantum friction threshold: the friction force is zero when the relative velocity is smaller than the speed of light within the slabs' materials. The origin of quantum friction was connected to the quantum Cherenkov radiation: the super-luminally moving object spontaneously emits photons. This concept is closely related to super-radiance, where a rotating body amplifies incident waves \cite{Zeldovich:1971:Gwrb}. In Refs.\,\cite{Farias:2017:Qfgs,Farias:2018:Qfttm}, quantum friction was calculated for two graphene sheets using the effective action approach, revealing a velocity threshold: the friction is zero when the relative velocity is smaller than the Fermi velocity. This correlation with Refs.\,\cite{Maghrebi:2013:SadCe,Maghrebi:2013:QCrnf} arises from the fact that the Dirac electron in graphene is described by the Dirac equation with the Fermi velocity instead of the speed of light. The threshold was confirmed by different calculations in Ref.\,\cite{Antezza:2023:Cfmg}. The nonperturbative approach was employed to study quantum friction in Ref.\,\cite{Brevik:2022:Feoe}, where the friction force was associated with electromagnetic instability: the kinetic energy of the relative motion transforms into exponentially growing coherent radiation.

In this study, we investigate the normal force between two laterally moving planes with isotropic conductivities. Previously \cite{Antezza:2023:Cfmg}, a general approach was developed for two conductive planes with relative lateral motion, which allowed for the calculation of both normal and tangential forces. The normal force was found to reduce to the usual Casimir force for planes with tensorial conductivities \cite{Fialkovsky:2018:Qcrcs}, with a specific form for the tensor of the moving plane being used. Quantum friction was also found to arise as an imaginary part of the energy calculated for complex frequencies, as discussed in earlier papers \cite{VanKampen:1968:mtVWf,Gerlach:1971:EvWFSSI,Silveirinha:2014:Oislepmm}.

In our paper, we focus specifically on the normal force for laterally moving planes with isotropic conductivities. As noted in previous studies \cite{Starke:2015:Faem,Starke:2016:RcOl}, Ohm's law for the moving plane needs to be considered carefully. The Lorentz transformation for a moving plane with scalar constant conductivity is not straightforward, as it involves a coefficient between $3$-vectors of the electric current and an electric field. The transformation law of the conductivity tensor has been discussed in a previous paper \cite{Starke:2016:RcOl}, utilizing a linear response tensor \cite{Melrose:2008:Qpup}. In the case of graphene, the role of linear response tensor plays the polarization tensor \cite{Bordag:2009:CipcgdDm,Antezza:2023:Cfmg}. To obtain the isotropic conductivity of a moving plane, we adopt the approach suggested in Ref.\,\cite{Antezza:2023:Cfmg} for graphene and consider the formal limit where the Fermi velocity and mass gap tend to zero. In this limit, the conductivity tensor becomes isotropic in the co-moving frame of the plane. However, in the laboratory frame, the conductivity is not diagonal and depends on the velocity. It is important to note that we employ the graphene approach as a computational tool only. The results obtained are applicable to various compositions with isotropic conductivity, including those with temporal dispersion (which was not considered in Ref.\,\cite{Maslovski:2011:Crmm}).

In previous studies \cite{Khusnutdinov:2014:Cescc,Khusnutdinov:2015:Cescp,Khusnutdinov:2016:Cescp,Khusnutdinov:2019:cei2dmss}, the Casimir and Casimir--Polder effects for planes with isotropic conductivity were explored. Two models of conductivity were employed: i) the constant conductivity, $\sigma = \sigma_0 \IB$, and ii) the Drude-Lorentz model with 7-oscillators $\sigma = \sigma_{\mathrm{DL}} (\omega) \IB$,

\begin{equation*}
	\sigma_{\mathrm{DL}} (\omega) = \frac{f_0 \omega_p^2}{\gamma_0 - \ii \omega} + \sum_{j=1}^7 \frac{\ii \omega f_j \omega_p^2}{\omega^2 - \omega_j^2 + \ii \omega \gamma_j}, 
\end{equation*}
with the parameters of this model obtained from experimental data for graphite from Ref.\,\cite{Djurisic:1999:Opg}. The first term in the model represents a Drude-like contribution, while the other terms have a Lorentz-like form. Since graphene is a single layer of graphite, the conductivity of graphene is obtained by multiplying the above expression by the interplane distance in graphite. The estimation of the binding energy per single sheet of graphene in graphite (a stack of graphenes) made in Ref.\,\cite{Khusnutdinov:2015:Cescp} revealed that the constant conductivity model underestimates the binding energy, whereas the Drude-Lorentz 7-oscillators model aligns well with experimental data.

Both models were employed to describe the Casimir effect in a stack of graphene layers and the Casimir-Polder effect for a micro-particle near the stack. The general case of a layered system consisting of conductive planes with tensorial conductivities was analyzed in Ref.\,\cite{Emelianova:2023:Cesgs}. This study demonstrated that the expressions for force and energy have the same form as those obtained for the case of scalar constant conductivity but with corresponding reflection coefficients for transverse electric (TE) and transverse magnetic (TM) modes.

In the case of constant conductivity, the Casimir energy exhibits a dependence of $1/a^3$ for all interplane distances. However, this relationship holds only for large interplane distances in the case of graphene, where the parameter $ma$ is much greater than one ($m$ is mass gap). This can be easily explained by noting that in the constant conductivity model, there are no dimensional parameters other than the interplane distance. For the case of small conductivity ($\eta_\gr = 2\pi \sigma_\gr  = 0.0114$ for graphene), the TM mode contributes linearly $\sim \eta$, while the TE mode contributes quadratically $\sim \eta^2$. However, this is not the case for two graphenes, where both modes contribute quadratically. As previously mentioned in Ref.\,\cite{Antezza:2023:Cfmg}, the spatial dispersion of conductivity plays an important role in the Casimir effect, causing the contribution to change from linear to quadratic.

The organization of this paper is as follows. In Sec.\,\ref{sec:1}, we re-derive the conductivity of a moving graphene sheet by applying boundary conditions and obtain the main formula for the normal Casimir energy. We discuss the problem of determining the eigenvalues of the product of reflection matrices. The Fresnel matrices do not commute, and their eigenvalues are not simply the product of the eigenvalues of the individual reflection matrices. Additionally, we briefly discuss the general property of lateral force along the planes and demonstrate that the necessary condition for this force is that the modulus of the Fresnel matrices is greater than one, indicating the production of photons. In Sec.\,\ref{sec:2}, we obtain the expressions for the normal Casimir force for isotropic conductivity in two scenarios: i) two identical planes and ii) an isotropic plane and a perfect metal. We perform numerical calculations of the Casimir energy and analytically derive the $v^2$ dependence for the velocity correction to the energy and pressure. We also evaluate the Drude-like model of isotropic conductivity numerically and demonstrate that, for large distances, the Drude-like model yields very similar results to the constant conductivity model. Finally, in the Conclusion, Sec.\,\ref{sec:3}, we discuss the results obtained in this study.

Throughout this paper, we utilize units where $\hbar = c = 1$.
 
\section{The Casimir energy of moving planes}\label{sec:1}

We use the approach for the Casimir effect of lateral moving graphene developed in Ref.\,\cite{Antezza:2023:Cfmg} as a computational trick. In order to provide a comprehensive understanding of this approach, we review the key steps of derivations of the normal force outlined in Ref.\,\cite{Antezza:2023:Cfmg} with some expanded explanations. 

The system under consideration involves two parallel conductive planes with isotropic conductivities and an inter-plane distance denoted as $a$. The first plane remains stationary in the laboratory frame, while the second plane undergoes lateral motion with a velocity $\vB$. The fluctuating electric field induces a current in the second plane following Ohm's law. This current affects the boundary condition and ultimately alters the energy spectrum. The current induced in the second conductive plane, which is in motion, is described by Ohm's law in its co-moving frame. To solve the scattering problem in the laboratory frame, it is necessary to determine the conductivity of the second plane in laboratory frame. The Lorentz transformation of Ohm's law was discussed in Ref.\,\cite{Starke:2015:Faem,Starke:2016:RcOl}. Given that Ohm's law has no a covariant form, as it represents a linear relationship between $3$-vectors of electric field and current density, the method of linear response tensor \cite{Melrose:2008:Qpup} is preferred in this particular case. Within the framework of this approach, the $4$-vector of current density $J^\mu$ and the 4-potential $A^\nu$ are linearly connected through a tensor of linear response $\Pi^\mu_\nu$: $J^\mu = \Pi^\mu_\nu A^\nu$. This is a covariant relation that can be Lorentz transformed into another inertial frame. This approach was successfully implemented in Ref.\,\cite{Starke:2015:Faem,Starke:2016:RcOl}, where it was demonstrated that the transformation of the conductivity tensor assumes a complex and non-standard form.

A similar methodology was applied in Ref.\,\cite{Antezza:2023:Cfmg} for a graphene sheet, where the polarization tensor serves as the linear response tensor. The complete action, which includes the Dirac electron, the classical electromagnetic field, and the effective action due to fermion loop correction \cite{Bordag:2009:CipcgdDm}, yields the following set of Maxwell equations

\begin{equation*}
	\partial_\mu F^{\mu\nu} = - \delta(z-a) \Pi^{\nu \alpha} A_\alpha = - 4\pi J^\nu, 
\end{equation*} 
where $z=a$ represents the position of the graphene plane and $\Pi^{\nu\alpha}$ denotes the polarization tensor resulting from the Dirac electron fermion loop. The current density assumes the form of a boundary condition. By integrating this relationship over an infinitesimally small interval $(a - \varepsilon, a + \varepsilon)$ with $\varepsilon$ approaching zero, it can be transformed into Ohm's law with the following conductivity tensor

\begin{equation*}
	\sigma^{ab} = \frac{\Pi^{ab}}{\ii \omega}.
\end{equation*}

The invariance of the boundary conditions with respect to 3-boosts

\begin{equation}\label{eq:Lorentz}
	\Lambda = 
	\begin{pmatrix}
		u^0 & - \uB\\
		-\uB & \IB + \frac{\uB \otimes \uB}{u^0 +1}
	\end{pmatrix}, \ \uB = (u^1,u^2),
\end{equation}
along the graphene plane was utilized in Ref.\,\cite{Antezza:2023:Cfmg} to determine the transformation of the conductivity tensor of graphene to the laboratory frame. Here, we employ this approach to calculate the boost transformation of the isotropic conductivity tensor. The conductivity tensor in the laboratory frame is represented by 

\begin{equation}\label{eq:eta2n}
	\sigmaB = \frac{\omega'}{\omega} \GB\sigmaB' \GB^T,
\end{equation}
where 

\begin{equation*}
	\GB =\IB  - \frac{\omega}{(ku)} \frac{\uB\otimes \uB}{u_0+1} + \frac{\uB \otimes \kB}{(ku)},
\end{equation*}
and $\omega', \sigmaB'$ are the frequency and the conductivity tensor in the co-moving frame, respectively.

In the framework of the scattering matrix approach \cite{Fialkovsky:2018:Qcrcs} the Casimir energy density per unit area $\mathcal{E}$ may be expressed in terms of the scattering matrix  $\bm{\mathcal{S}}$ 

\begin{equation}\label{eq:SGen}
	\mathcal{E} = \frac{\ii}{4\pi} \iint \frac{\dd^2k}{(2\pi)^2} \int_0^\infty \ln\det \bm{\mathcal{S}} \frac{k_3 \dd k_3}{\sqrt{\kB^2 + k_3^2}},
\end{equation}
where the scattering matrix of the total system consists of the reflection $\RB$ and transmission $\TBB$ matrices

\begin{equation*}
	\bm{\mathcal{S}} = 
	\begin{pmatrix}
		\RB & \TBB' \\
		\TBB & \RB'
	\end{pmatrix}.
\end{equation*}

This matrix describes the scattering of the electromagnetic field 

\begin{equation*}
	\begin{pmatrix}
		\Vl{\Eb}_l\\
		\Vr{\Eb}_r
	\end{pmatrix} =
	\bm{\mathcal{S}} \begin{pmatrix}
		\Vr{\Eb}_l\\
		\Vl{\Eb}_r
	\end{pmatrix},
\end{equation*}
where the indexes $l$ and $r$ stand for electromagnetic field on the left and right sides of the total system, correspondingly.  The system, in general,  may consist of a set of planes.  The vectors over the fields denote the wave direction. The scattering matrix  $\bm{\mathcal{S}}$ can not be reduced to a product of matrices for each plane \cite{Emelianova:2023:Cesgs}.  

The general relation may be transformed \cite{Fialkovsky:2018:Qcrcs} to the following expressions for the energy density $\mathcal{E} $ and pressure $\mathcal{P}$ for real frequencies

\begin{equation}\label{eq:Efirst}
	\mathcal{E}  = - \frac{1}{2 \ii } \iint \frac{\dd^2 k}{(2\pi)^3}\left( I_- - I_+\right),\ \mathcal{P}  = \iint \frac{\dd^2 k}{(2\pi)^3}\left( J_- +  J_+\right),
\end{equation}
where 

\begin{align}
	I_\pm &= \int_k^\infty \dd \omega \ln \det  \left[ 1 - e^{ \pm 2 \ii a k_3} \RC(\pm k_3) \right],\nonumber\\ 
	J_\pm &= \int_k^\infty \dd \omega k_3 \frac{e^{ \pm 2 \ii a k_3} (\tr \RC(\pm k_3)  - 2e^{ \pm 2\ii a k_3} \det \RC(\pm k_3))}{\det  \left[ \IB - e^{ \pm 2 \ii a k_3} \RC(\pm k_3) \right]},\label{eq:IJ}
\end{align}
$\RC(\pm k_3) = \rB'_1(\pm k_3) \rB_2(\pm k_3)$, and $k_3 = \sqrt{\omega^2 - \kB^2}$. These formulas consider only the propagating waves, because $\omega \geq k$. The subscript $1(2)$ means that all reflection matrices related to the rest (moving) plane. The scattering matrix for each part of system (each plane) has the following form 

\begin{equation*}
	\bm{\mathcal{S}} = 
	\begin{pmatrix}
		\rB & \tB' \\
		\tB & \rB'
	\end{pmatrix},
\end{equation*}
with corresponding index and argument.  It connects the electromagnetic waves on the left side $(l)$ of the specific plane of the system with that on the right $(r)$ side by relation 

\begin{equation*}
	\begin{pmatrix}
		\Vl{\Eb}_l\\
		\Vr{\Eb}_r
	\end{pmatrix} =
	\bm{\mathcal{S}} \begin{pmatrix}
		\Vr{\Eb}_l\\
		\Vl{\Eb}_r
	\end{pmatrix},
\end{equation*}
where the direction of the vector indicates the direction of the wave, $\Vr{\Eb} \sim e^{+\ii k_3 z}$, $\Vl{\Eb} \sim e^{-\ii k_3 z}$.

The reflection matrices for the conductive plane were derived in Ref.\,\cite{Fialkovsky:2018:Qcrcs} by using the boundary condition on the plane

\begin{equation}
	\rB_i = \rB' _i = - \frac{\omega^{2} \etaB_i - \kB \otimes (\kB\etaB_i) + \IB \omega k_3 \det\etaB_i}{\omega^2 \tr \etaB_i - \kB\kB\etaB_i + \omega k_3 (1 + \det \etaB_i)},\label{eq:r}
\end{equation}
where $\etaB_i = 2\pi \sigmaB_i$, and $\sigmaB_i$ is the conductivity tensor of the plane $i=1,2$ ($i=1$ is at rest and $i=2$ is moving in laboratory frame).  In domain $\omega <k$ there are evanescent and waveguide modes \cite{Bordag:2012:Evetpstswpm}, but as demonstrated in this paper, by rotation of the contour of integration to the imaginary axis, the contribution of these modes are cancelled out with energy of boundary states of corresponding modes. 

\begin{figure}
	\centering
	\includegraphics[width=0.5\linewidth]{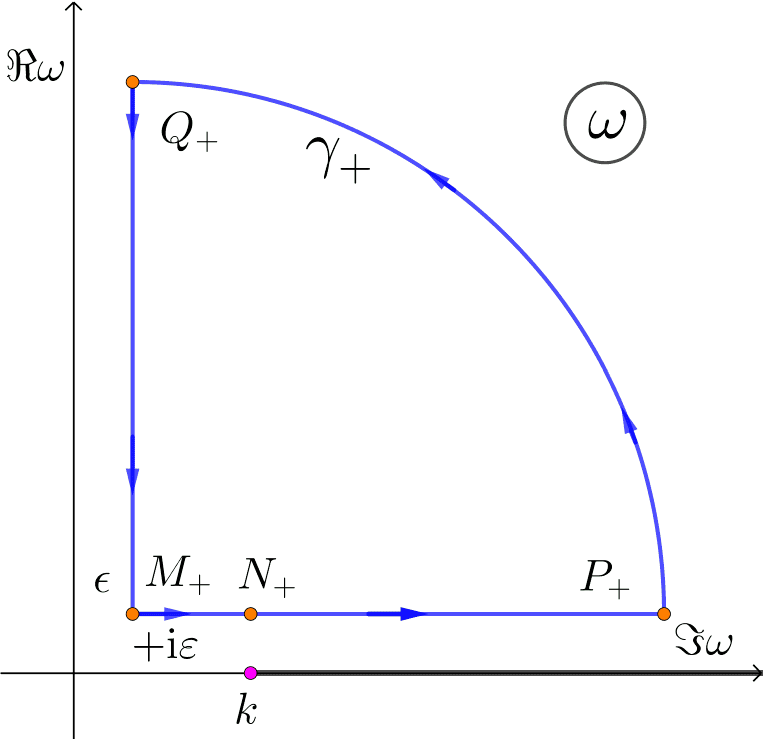}\includegraphics[width=0.5\linewidth]{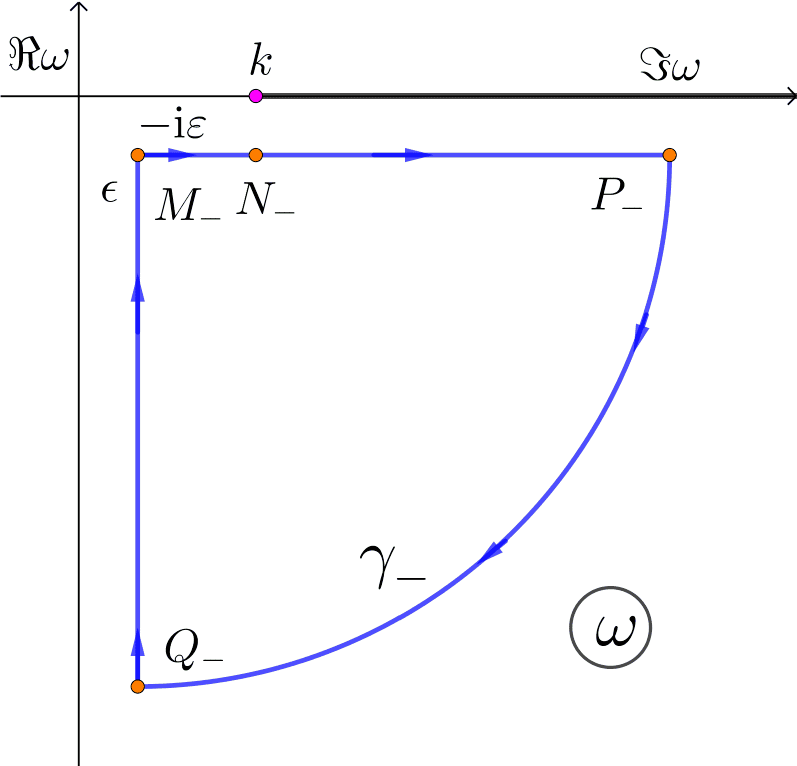}
	\caption{The contour of integration for $I_\pm$ and $J_\pm$ in Eq.\,\eqref{eq:IJ} is depicted. The integration over imaginary frequency $\omega = \ii \xi$ yields the energy $\mathcal{E}^\perp$. The potential presence of poles within these contours may contribute to a non-zero energy $\mathcal{E}^\parallel$.}
	\label{fig:path}
\end{figure}

Then we make a rotation of the integration contour over $\Re\,\omega$ in $I_\pm$ and $J_\pm$ \eqref{eq:IJ} to the imaginary axis (see Fig.\,\eqref{fig:path}).  

After the rotation of the contour to the imaginary axis, two contributions survive, which we refer to as $\mathcal{E}^\perp$ and $\mathcal{E}_\parallel$, corresponding to the normal (real) and parallel (imaginary) to the planes, respectively. The first contribution with integration along the imaginary frequency $\omega = \ii \xi$ is given by

\begin{equation}
	\mathcal{E}^\perp  =  \iint \frac{\dd^2 k}{2(2\pi)^3} \int_{-\infty}^{+\infty} \dd\xi  \ln \det  \left[ 1 - e^{ -2 a k_E}  \RC(\ii k_E) \right], \label{eq:E2}
\end{equation}
where $k_E = \sqrt{\xi^2 + k^2}$. The corresponding force is perpendicular to the planes, as in the usual Casimir force. This expression may be simplified by using eigenvalues of matrix $\RC$ and be represented as a sum of TE and TM contributions. The matrices $\rB'_1$ and $\rB_2$ do not commute (see below), and therefore the eigenvalues of $\RC$ are not a product of the eigenvalues of $\rB'_1$ and $\rB_2$. The eigenvalues of $\RC$ may be found in closed, albeit complicated, forms \cite{Antezza:2023:Cfmg}, which correspond to the contributions of TM and TE modes separately. Instead of this approach, we use the expression for energy in the form \cite{Fialkovsky:2018:Qcrcs} directly via the conductivity matrices $\etaB_i$

\begin{align}
	\mathcal{E}^\perp &= \int \frac{\dd^2 k}{2(2\pi)^3} \int_{-\infty}^\infty \dd\xi \ln \left(1 + e^{-4 a k_E} \frac{\xi^2 k_E^2 }{b_1b_2} \det \etaB_1 \det \etaB_2 \right. \nonumber\\ 
	&-\left.  e^{-2 a k_E}\left[\frac{\xi^2 k_E^2 }{b_1b_2} \left[(1- \det\etaB_1) (1- \det\etaB_2) + \det(\etaB_1 - \etaB_2) \right] - \frac{\xi k_E}{b_1}   - \frac{\xi k_E}{b_2}   + 1 \right] \right),\label{eq:XY}
\end{align}
where $b_i = \xi^2 \tr \etaB_i + (\kB\kB\etaB_i)  + \xi k_E \bigl(1+ \det\etaB_i\bigr)$. The second contribution $\mathcal{E}_\parallel$ contributes to the force along the planes, representing the Casimir friction. 

If the tensor conductivity has the following structure  

\begin{equation}\label{eq:etaBi}
	\etaB_i = \eta_i^\te \IB + (\eta_i^\tm - \eta_i^\te) \frac{\kB \otimes \kB}{\kB^2},
\end{equation}
it possesses eigenvalues $\eta_i^\tm,\eta_i^\te$ which corresponds to TM and TE modes. In particular, the graphene conductivity tensor has such a structure \cite{Bordag:2009:CipcgdDm} with corresponding conductivities of modes. The Hall conductivity gives an additional antisymmetric term \cite{Antezza:2020:CftamccpteC}. In this case, the expression \eqref{eq:XY} for Casimir energy may be transformed to the well-known form of sum contributions of the TE and TM modes

\begin{equation*}
	\mathcal{E}^\perp = \int \frac{\dd^2 k}{2(2\pi)^3} \int_{-\infty}^\infty \dd\xi \left[\ln \left(1 - \frac{e^{-2 a k_E}}{\left(1 + \frac{k_E}{\xi \eta_1^\te}\right) \left(1 + \frac{k_E}{\xi \eta_2^\te}\right)}\right)  + \ln \left(1 - \frac{e^{-2 a k_E}}{\left(1 + \frac{k_E}{\xi \eta_1^\tm}\right) \left(1 + \frac{k_E}{\xi \eta_2^\tm}\right)}\right)\right].
\end{equation*}
In the case under consideration, the conductivity of the moving plane does not have form \eqref{eq:etaBi}, and we use the Casimir energy in the form \eqref{eq:XY}.  
 
In Ref.\,\cite{Antezza:2023:Cfmg}, the normal force was considered for moving graphene with the following conductivity in a co-moving frame 

\begin{equation}\label{eq:eta1}
	\etaB_1 = \eta_{\gr} \frac{\kT}{\omega} \left(\IB + v_F^2 \frac{\kB \otimes \kB}{\kT^2}\right) \PhiT \left(\frac{\kT}{2m}\right),
\end{equation}
where $\eta_\gr = 2\pi \sigma_\gr = \pi e^2/2$, $v_F$ is the Fermi velocity, and 

\begin{equation}
	\PhiT(y) = \frac{2\ii }{\pi y}\left\{ 1 - \frac{y^2 + 1}{y}\arctanh y\right\},\ \kT =  \sqrt{\omega^2 - v_F^2 \kB^2}. \label{eq:kT}
\end{equation}

In the laboratory frame, the general structure of the moving plane's conductivity tensor \eqref{eq:eta2n} takes the form \cite{Antezza:2023:Cfmg}

\begin{equation}
	\etaB_2 = i_1 \IB  + i_2 \kB\otimes\kB + i_3 (\kB\otimes\vB + \vB\otimes\kB),
\end{equation}
where

\begin{align}
	i_1 &= \frac{\eta _\gr \PhiT'}{\omega  \kB^2 \kT'}  \left(\kB^2 \kT'^2 + \frac{1-v_F^2}{1-v^2} ((\kB\vB)^2 - \kB^2 \vB^2)k_3^2\right),\nonumber\\ 
	i_2 &= \frac{\eta _\gr \PhiT'} {\omega \kB^2 \kT'} \left(\kB^2 v_F^2 +\frac{1-v_F^2}{1-v^2} \vB^2 k_3^2 \right),\nonumber\\ 
	i_3 &= \frac{\eta_\gr \PhiT'}{\kB^2 \kT'}  \frac{1-v_F^2}{1-v^2} \left(\kB^2 - \omega(\kB\vB)\right),
\end{align}
and

\begin{equation}\label{eq:ktp}
	\kT' = \sqrt{\kT^2 + \frac{1-v_F^2}{1-v^2} (\omega^2 \vB^2 + (\kB\vB)^2 -2 \omega (\kB\vB) )},
\end{equation}
is the Lorentz transformation of $\kT$ \eqref{eq:kT}. 

The reflection matrices  $\rB_i$ \eqref{eq:r}, on the basis of eigenvectors, are given as

\begin{equation}
	\hat{\rB}_1(k_3) = - 
	\begin{pmatrix}
		\frac{k_3 \eta}{\eta k_3 + \kT} & 0 \\
		0& \frac{\eta \kT}{\eta \kT + k_3}
	\end{pmatrix}, \hat{\rB}_2(k_3) = - 
	\begin{pmatrix}
		\frac{k_3 \eta '}{\eta' k_3 + \kT'} & 0 \\
		0& \frac{\eta' \kT'}{\eta' \kT' + k_3}
	\end{pmatrix}, \label{eq:eigenr1andt2}
\end{equation}
where $\eta = \eta_\gr \PhiT(y)$,  $\eta' = \eta_\gr \PhiT(y')$, and

\begin{equation}
	\TB_i^{-1} \rB_i \TB_i = \hat{\rB}_i \Leftrightarrow \rB_i  = \TB_i \hat{\rB}_i\TB_i^{-1} .
\end{equation}
The matrices $\TB_i $ diagonalize the reflection matrices $\rB_i$ and have the form 

\begin{equation*}
	\TB_1 = 
	\begin{pmatrix}
		k_2&k_1 \\
		-k_1&k_2
	\end{pmatrix}, 
	\TB_2 = 
	\begin{pmatrix}
		k_2-v_2\frac{k_3^2 }{\omega -(\kB\vB)}  & k_1 - v_1 \omega \\[1ex]
		-k_1+ v_1\frac{k_3^2}{\omega -(\kB\vB)} &k_2- v_2 \omega
	\end{pmatrix}. 
\end{equation*} 
It is worth noting that the eigenvalues of $\rB_2$ can be obtained from the eigenvalues of $\rB_1$ through a Lorentz transformation. Specifically, $k_3 \to k_3$,  $\kT \to \kT'$, and $\eta \to \eta'$ under these transformations, resulting in $\hat{\rB}_1 \to 	\hat{\rB}_2$.  This is expected as eigenvalues are invariants of a matrix. 

The eigenvector basis of $\rB_1$,

\begin{equation*}
	\bm{a}_1 = (k_2,-k_1),\  \bm{a}_2 = (k_1,k_2) = \kB\label{eq:basisr1}
\end{equation*}
is orthogonal with $\bm{a}_1\cdot\bm{a}_2 = 0$, and $\bm{a}_1^2 = \bm{a}_2^2 = \kB^2$. On the other hand, the eigenvector basis of $\rB_2$,

\begin{align*}
	\bm{c}_1 &= \left(k_2- v_2\frac{k_3^2 }{ \omega -(\kB\vB)}, -k_1+ v_1\frac{k_3^2 }{ \omega -(\kB\vB)}\right),\nonumber\\ 
	\bm{c}_2 &= \left(k_1 - v_1 \omega, k_2- v_2 \omega\right),\label{eq:basisr2}
\end{align*}
is not orthogonal:

\begin{equation}
	\bm{c}_1 \cdot \bm{c}_2 = \frac{\kB^2 -\omega (\kB\vB)}{\omega -(\kB\vB)}(\kB\vB\nB) \not = 0,
\end{equation}
where $n^a = \delta^a_3$.

Through straightforward calculations, the following expression for the commutator

\begin{equation}
	[\rB_1, \rB_2]  =\eta \eta'  \frac{(1- v_F^2)^2}{1-v^2} \frac{\omega  ((\kB\vB) \omega - k^2)}{\kT\kT' b_1 b_2}(k_3^2 \kB \otimes \vB  - \omega ^2 \vB\otimes \kB), \label{eq:commr12}
\end{equation}
is obtained. Therefore, the commutator $[\rB_1, \rB_2] \not = 0$ and the $\rB_1$ and $\rB_2$ can not be diagonalized simultaneously. It means that the eigenvalues of $\rB_1\cdot \rB_2$ are not a product of the eigenvalues of matrices $\rB_1$ and $\rB_2$. 

The straightforward calculations give the following expression for eigenvalues of matrix $\RC$:

\begin{align}
	r_{\tm/\te}(k_3)&= \frac{\eta  \eta '}{2 P QP'  Q'} \left\{ \alpha  \left(k_3 \left(k^2 Q'+Q \left((ku)^2-k_3^2\right)\right)+\alpha  \left((ku) \omega -k_3^2 u_0\right){}^2\right) +2 k_3^2 Q Q'  \right. \nonumber \\
	&\left. \pm \alpha \sqrt{\left(k_3 \left(k^2 Q'+Q \left((ku)^2-k_3^2\right)\right)+ \alpha  \left((ku) \omega -k_3^2 	u_0\right){}^2\right)^2 + 4 k_3^4 Q Q' \left((\kB \uB)^2-k^2 u^2\right)}\right\},\label{eq:eigenR}
\end{align}
where, $\alpha = 1- v_F^2$, and 

\begin{equation*}
Q = \kT \eta +k_3, P = \kT + \eta k_3, Q' = \kT' \eta' +k_3, P' = \kT' + \eta' k_3.
\end{equation*}
In the case of zero velocity, $\uB=\bm{0}$, $\hat{\rB}  = \hat{\rB}_1^2$, is obtained, as expected. To establish the correct correspondence, the following square root sign convention

\begin{equation}
\sqrt{\left(\alpha  k^2+2 k_3 Q\right)^2} = \alpha  k^2+2 k_3 Q,
\end{equation}
is employed. The eigenvalues \eqref{eq:eigenR} have a highly complex form. Hence, expression \eqref{eq:XY} is used where the Casimir energy is directly expressed in terms of the tensors' conductivities.

Let us briefly discuss the second contribution $\mathcal{E}_\parallel$ which may contribute to the force along planes. The contribution to the Casimir pressure takes the form

\begin{align}
	\mathcal{P}^\parallel  &=\iint \frac{\dd^2 k}{(2\pi)^3}\left\{  \oint_{\gamma_-} \dd \omega k_3 \frac{e^{ - 2 \ii a k_3} (\tr \RC(- k_3)  - 2e^{ - 2\ii a k_3} \det \RC(- k_3))}{\det  \left[ \IB - e^{ - 2 \ii a k_3} \RC(- k_3) \right]}\right.\nonumber \\
	&+\left. \oint_{\gamma_+} \dd \omega k_3 \frac{e^{2 \ii a k_3} (\tr \RC(k_3) - 2e^{2 \ii a k_3} \det \RC(k_3))}{\det  \left[ \IB - e^{2 \ii a k_3} \RC(k_3) \right]}\right\},
\end{align}
with the contours $\gamma_\pm$ depicted in Fig.\,\ref{fig:path}. This expression has a non-zero value only if poles appear inside the contours, satisfying the relations

\begin{equation}\label{eq:gencond}	
	\det  \left[ \IB - e^{ \pm 2 \ii a k_3} \rB'_1(\pm k_3) \rB_2(\pm k_3) \right] = 0.
\end{equation}
This relation, with various contexts, has been noted in previous works  \cite{Gerlach:1971:EvWFSSI,VanKampen:1968:mtVWf,Henkel:2004:CspCf,Maghrebi:2013:QCrnf,Silveirinha:2014:Oislepmm}. If the matrix $\RC$ has eigenvalues $r_\tm$ and $r_\te$, this relation can be separated into two scalar relations

\begin{equation}\label{eq:fricGen}
	1 - e^{ \pm 2 \ii a k_3} r_\tm(\pm k_3)= 0,\ 1 - e^{ \pm 2 \ii a k_3} r_\te (\pm k_3)= 0.
\end{equation}
The solutions of these relations must possess imaginary parts to contribute as residues.

It can be demonstrated that solutions to these relations exist if and only if 

\begin{equation}\label{eq:fricRes}
	\left|r_\tm(\pm k_3)\right| >1, \left|r_\te(\pm k_3)\right| >1.
\end{equation}
Since $\sign(\Im k_3) = \sign (\Im \omega)$, then 

\begin{equation*}
	\left| e^{\pm 2\ii a k_3}\right| = \left[e^{\pm 2\ii a k_3} e^{\mp 2\ii a k_3^*}\right]^{1/2} = e^{\mp 2a \Im k_3} = e^{- 2a |\Im k_3|} <1.  
\end{equation*}
Taking this inequality into account, the relations \eqref{eq:fricRes} can be obtained from \eqref{eq:fricGen}.  

Therefore, the condition $\mathcal{P}^\parallel \not = 0$ is associated with virtual photon production because the modulus of the reflection coefficients is greater than unity. Lifshitz demonstrated in Ref.\,\cite{Lifshitz:1956:tmafbs} that without relative velocity Eq.\,\eqref{eq:gencond} has no solutions because the inequalities  \eqref{eq:fricRes} can not be satisfied. This statement can be proven for graphene with zero mass gap (for simplicity). For zero velocity $\vB = 0$ and $m=0$ 

\begin{align}
	r_\tm^{-1}(\pm k_3) &= \left(1 \pm \frac{\kT}{\eta k_3}\right)^2 =  \left(1 + \frac{\kT}{\eta_\gr k_3}\right)^2, \nonumber \\
	r_\te^{-1}(\pm k_3) &= \left(1 \pm \frac{k_3}{\eta \kT}\right)^2 = \left(1 + \frac{k_3}{\eta_\gr \kT}\right)^2, \label{eq:rtmte}
\end{align}
is obtained. Then, 

\begin{equation}
	\left| r_\tm^{-1}(\pm k_3) \right| = \left(1 + \frac{|\kT|^2}{\eta^2_\gr |k_3|^2 } + 2\frac{\Re \kT \Re k_3 + \Im \kT \Im k_3}{\eta_\gr |k_3|^2 }\right)^2.
\end{equation}
It can be easily shown that $\Re \kT \Re k_3 + \Im \kT \Im k_3 > 0$ thus satisfying 

\begin{equation}
	\left| r_\tm^{-1}(\pm k_3) \right|  > 1.
\end{equation}
Hence, we conclude that the relations \eqref{eq:fricRes} cannot be satisfied in this case, indicating the absence of solutions for Eq.\,\eqref{eq:gencond}. However, it is expected that solutions can arise due to the relative motion of the planes.

\section{The case of an isotropic conductivity}\label{sec:2}

The case of isotropic conductivity can be obtained by taking the formal limits $v_F\to 0$ and $\PhiT\to 1\ (m\to 0)$ in Eq.\,\eqref{eq:eta1} and changing $\eta_\gr$ to conductivity of corresponding plane $\eta_i$. After taking these limits, the conductivity tensor for the plane at rest becomes diagonal $\etaB_1 = \eta_1 \IB$ and $\kT'  = \gamma \omega_v$, where $\omega_v = \omega - \kB\vB$, $\gamma =1/\sqrt{1-v^2}$ being the relativistic factor, and  

\bs\label{eq:eta2}
\begin{equation}
	\etaB_2 = i'_1 \IB  + i'_2 \kB\otimes\kB + i'_3 (\kB\otimes\vB + \vB\otimes\kB),
\end{equation}
where 

\begin{equation}
	i'_1 = \frac{\eta_2 \gamma \left(\kB^2 \omega_v^2 + ((\kB\vB)^2 - \kB^2 \vB^2)k_3^2\right)}{ \kB^2 \omega \omega_v} ,\ 	i'_2 = \frac{\eta_2 \gamma \vB^2 k_3^2} {\kB^2 \omega  \omega_v},\ i'_3 = \frac{\eta_2 \gamma }{\kB^2 \omega_v}  (\kB^2 - \omega(\kB\vB)).
\end{equation}
\es
The quantity $\omega_v \gamma$ represents the frequency of photons in a laboratory that was emitted in a co-moving frame.

Performing straightforward calculations at the imaginary axis $\omega = \ii \xi$ we obtain

\begin{align}
	\ b_1 &= \left(\eta_1  \xi + k_E\right) \left(\xi + \eta_1 k_E\right),\ b_2 =\frac{\xi}{\gamma\xi_v} \left(\eta_2 \gamma \xi_v + k_E\right) \left(\gamma \xi_v + \eta_2  k_E \right),\nonumber\\ 
	\det(\etaB_1-\etaB_2) &=  (\eta_1-\eta_2)^2 + \eta_1\eta_2\left(\gamma v^2\frac{ k_E^2 }{\xi  \xi_v} + 2 (1-\gamma) \right),\ \det\etaB_i = \eta_i^2,  
\end{align} 
where $\xi_v = \xi + \ii \kB\vB$. Then we use the polar coordinates for $\kB$ in Eq.\,\eqref{eq:XY}, $\kB\vB = k v \cos\varphi$ and  transform the coordinates of the plane  $k \in [0,\infty)$, $\xi \in (-\infty,\infty)$ to polar coordinates $\xi = k_E \cos\theta, k = k_E \sin\theta$. After these changes, the dependence of $k_E$ only survives in the exponents. By changing the variable $a k_E = y$ we observe that the energy depends on the inter-plane distance as $1/a^3$ for constant conductivities, as expected \cite{Khusnutdinov:2014:Cescc}. Thus, the energy and pressure have the following form ($x=\cos\theta$)

\begin{equation}\label{eq:Casimir}
	\mathcal{E}^\perp_{1,2} = \Re \int_0^\infty  \frac{y^2\dd y}{(2\pi a)^3} \int_0^1 \dd x \int_0^\pi \dd \varphi E_{1,2}, \ \mathcal{P}^\perp_{1,2}  = \frac{3}{a}\mathcal{E}^\perp_{1,2},  
\end{equation} 
where 

\begin{align}
	E_{1,2}&= \ln \left(1 + e^{-4 y} \frac{x^2\eta_1^2\eta_2^2}{\beta_1\beta_2}\right.\nonumber\\ 
	  &\left.- e^{-2 y}\left(\frac{x^2}{\beta_1\beta_2} \left[(1- \eta_1\eta_2)^2 + \eta_1\eta_2\left(\frac{\gamma v^2 }{x x_v} + 2 (1-\gamma) \right) \right] - \frac{x}{\beta_1}   - \frac{x}{\beta_2}   + 1 \right)\right), \label{eq:E12}
\end{align}
and

\begin{equation}
	\beta_1 = \left(\eta_1  x + 1\right) \left(x + \eta_1 \right),\ \beta_2 =\frac{x}{\gamma x_v} \left(\eta_2 \gamma x_v + 1\right) \left(\gamma x_v + \eta_2 \right),\ x_v = x + \ii v \sqrt{1-x^2} \cos\varphi.
\end{equation}

If the first plane (at rest) is a perfect conductor, we take the limit $\eta_1 \to \infty$ and obtain

\begin{equation}\label{eq:Eid2}
	E_{\id,2}= \ln \left(1 + e^{-4 y} \frac{x\eta_2^2}{\beta_2} - e^{-2 y}\left(\frac{x \eta_2^2}{\beta_2}   - \frac{x}{\beta_2}   + 1 \right)\right), 
\end{equation}
whereas for two ideal planes,

\begin{equation}
	E_{\id,\id}= 2\ln \left(1 + e^{-2 y}\right), 
\end{equation}
the energy does not depend on the velocity.

The expressions \eqref{eq:Casimir}, and \eqref{eq:E12} coincide with those obtained in Ref.\,\cite{Philbin:2009:Nqfump}, where Philbin and Leonhardt considered two plates of finite thickness  with relative lateral motion. As noted in Ref.\,\cite{Bordag:2006:LfgscnvWCi}, the usual reflection coefficients can not be used for 2D materials due to the impossibility of taking the limit of zero thickness. The reflection coefficients in this case have to be calculated  using scattering theory \cite{Jaekel:1991:Cfptm} or 2D Quantum Electrodynamics \cite{Bordag:2009:CipcgdDm}. In the case under consideration, we have to use the reflection coefficients \eqref{eq:rtmte} for the plane at rest and the same expressions, but with a boosted wave vector \eqref{eq:ktp} for the moving plane. Then, one takes limit $v_F \to 0$ to obtain isotropic conductivity. Finally, we arrive at \eqref{eq:Casimir} and \eqref{eq:E12}. For example, the coefficient at $e^{-4y}$ in \eqref{eq:E12}

\begin{equation*}
	\frac{x^2\eta_1^2\eta_2^2}{\beta_1\beta_2} = r_{E1}r_{E2}r_{B1}r_{B2}
\end{equation*}
in notations of Ref.\,\cite{Philbin:2009:Nqfump}. 

Without a relative movement, $\vB = 0$, we return to the results obtained in Ref.\,\cite{Khusnutdinov:2014:Cescc}:

\begin{equation}
	E= \ln \left(1- e^{-2y} \frac{\eta_1 \eta_2}{(\eta_1 +x)(\eta_2 +x)}\right) + \ln \left(1- e^{-2y} \frac{\eta_1\eta_2 x^2}{(x\eta_1 +1)(x\eta_2 +1)}\right) = E_\tm + E_\te,
\end{equation}
which is the sum of TM and TE contributions. 

\begin{figure}
	\centering
	\includegraphics[width=0.33\linewidth]{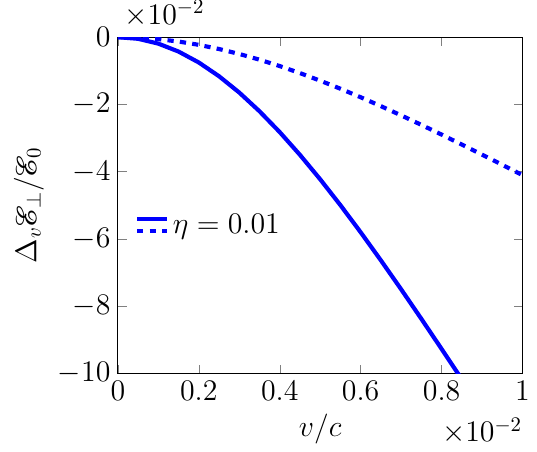}\includegraphics[width=0.33\linewidth]{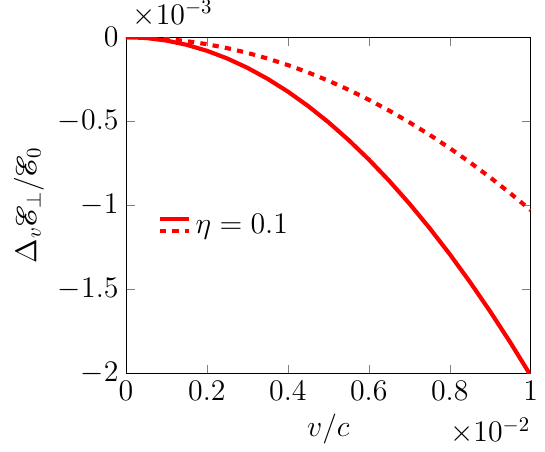}%
	\includegraphics[width=0.33\linewidth]{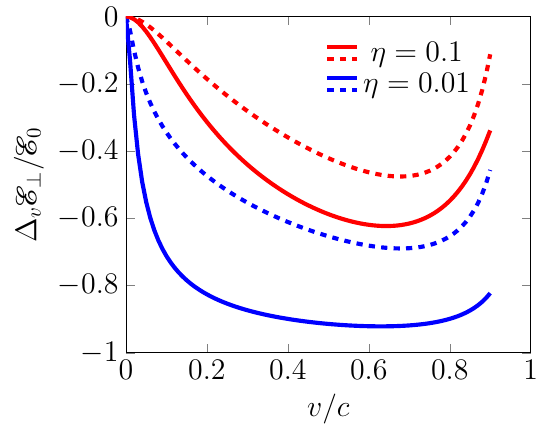} 
	\caption{The plots illustrating the velocity correction $\frac{\Delta_v \mathcal{E}_\perp}{\mathcal{E}_0} = \frac{ \mathcal{E}_\perp - \mathcal{E}_0}{\mathcal{E}_0}$ for $\eta= 0.01, 0.1$ are presented. In the case of small velocity and conductivity, where $v\ll \eta \ll 1$ (shown in the left and middle panels), a quadratic correction is observed in accordance with Eq.\,\eqref{eq:v2}. The solid lines represent two planes with equal conductivity as given by Eq.\,\eqref{eq:E12}, while the dashed lines represent systems where one plane is stationary with perfect conductivity (described by Eq.\,\eqref{eq:Eid2}), and the second plane possesses isotropic conductivity (described by Eq.\,\eqref{eq:E12}).}\label{fig:en}
\end{figure}

Let us consider the constant conductivities case with equal conductivities: $\eta_1 = \eta_2 = \eta = const$. For $v=0$ we obtain from Eq.\,\eqref{eq:Casimir}  the sum of TM and TE modes contribution, denoted as

\begin{align}
		E_{1,2}^0&= \ln \left(1 - e^{-2 y}\frac{\eta^2}{(x+\eta)^2}\right) + \ln \left(1 - e^{-2 y}\frac{x^2\eta^2}{(1+x \eta)^2}\right), \nonumber\\ 
		E_{\id,2}^0&= \ln \left(1 - e^{-2 y}\frac{\eta}{x+\eta}\right) + \ln \left(1 - e^{-2 y}\frac{x\eta}{1+x \eta}\right).
\end{align}
In the case $\eta \ll 1$ the Casimir energy for $v=0$ has the following form 

\begin{align}
	\mathcal{E}^0_{1,2}& = \frac{\eta}{a^3}\frac{180 \zeta_R (3) + \pi^4 + 60 \pi^2 - 1440 \ln 2}{2880 \pi^2} = -3.2\cdot 10^{-3} \frac{\eta}{a^3},\nonumber\\ 
	\mathcal{E}^0_{\id,2}& = \frac{\eta}{a^3} \left[\frac{\ln\eta}{32 \pi^2} + \frac{90 \zeta_R (3) + \pi^4 + 15 \pi^2 - 405}{2880 \pi^2}\right] = -  \left[3.1 \ln\eta^{-1} + 1.8\right] \cdot 10^{-3}\frac{\eta}{a^3}.
\end{align}
In SI dimensional units, we have to multiply the above relations by $\hbar c = 3.16\cdot 10^{-26}$ J m.

For small velocity and conductivity, specifically when $v \ll \eta \ll 1$ we obtain from Eq.\,\eqref{eq:Casimir} the expressions

\begin{equation}\label{eq:v2}
	\frac{\Delta_v\mathcal{E}^\perp_{1,2}}{\mathcal{E}^0_{1,2}} = \frac{\Delta_v\mathcal{P}^\perp_{1,2}}{\mathcal{P}^0_{1,2}} \approx -0.19 \left(\frac{v}{\eta}\right)^2, \frac{\Delta_v\mathcal{E}^\perp_{\id,2}}{\mathcal{E}^0_{\id,2}} = \frac{\Delta_v\mathcal{P}^\perp_{\id,2}}{\mathcal{P}^0_{\id,2}} \approx -\frac{0.3}{\ln\eta^{-1} + 0.57}\left(\frac{v}{\eta}\right)^2, 
\end{equation}
where $\Delta_v \mathcal{E}^\perp_{ik} = \mathcal{E}^\perp_{ik} - \mathcal{E}^0_{ik}$. The relative velocity correction is quadratic in the velocity and is negative meaning the force is decreased due to the motion of the planes.

Numerical evaluations of \eqref{eq:Casimir}  are shown in Fig.\,\ref{fig:en} for two systems: $(1,2)$ -- two conductive planes with constant conductivity $\eta$ (solid lines) and $(\id,2)$ -- the first plane at rest is a perfect metal (dashed lines). We calculated velocity correction to the energy:  $\Delta_v \mathcal{E}_\perp/\mathcal{E}_0 = (\mathcal{E}_\perp - \mathcal{E}_0)/\mathcal{E}_0$, where $\mathcal{E}_0$ is energy without relative movement. The velocity correction is negative for both systems and exhibits  quadratic behaviour $\sim v^2$ for small velocity $v \ll \eta \ll 1$. The absolute value of correction is greater for the first system. 

The above derivation is applicable for isotropic but frequency-dependent (temporal dispersion) conductivity, $\eta = \eta (\omega)$. Let us consider the simple case of Drude-like conductivity with 

\begin{equation}\label{eq:Drude}
	\eta_1 = \frac{\eta \Gamma}{\Gamma + \xi}\ ,\ \eta_2 = \frac{\eta \Gamma}{\Gamma + \gamma\xi_v},
\end{equation}
where parameters are taken for graphene, $\Gamma = 6.365$ eV and $\eta = \eta_\gr = e^2/4$ \cite{Khusnutdinov:2015:Cescp}. After changing the integrand variables as described above, the Casimir energy acquires an additional dependence on the inter-plane distance through conductivity:

\begin{equation}
	\eta_1 = \frac{\eta_\gr (a\Gamma)}{(a\Gamma) + y x}\ ,\ \eta_2 = \frac{\eta_\gr (a\Gamma)}{(a\Gamma) + \gamma y x_v}.
\end{equation} 
For $a=100$ nm and  $\Gamma = 6.365$ eV, one has, $a\Gamma = 3.225$. For large values of $a\Gamma \gg 1$ the conductivities $\eta_1 = \eta_2 = \eta_\gr$, consistent with the constant conductivity model, are valid for large inter-plane distances.  
\begin{figure}
	\centering
	\includegraphics[width=0.5\linewidth]{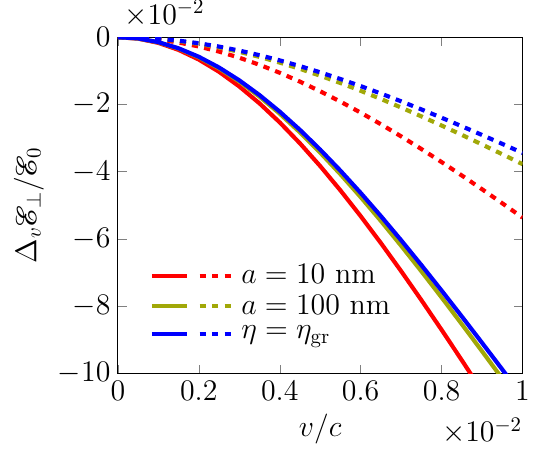}\includegraphics[width=0.5\linewidth]{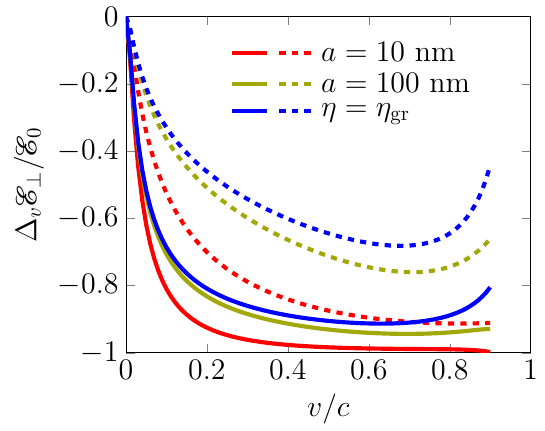}
	\caption{The plots display the velocity correction $\frac{\Delta_v \mathcal{E}_\perp}{\mathcal{E}_0} = \frac{ \mathcal{E}_\perp - \mathcal{E}_0}{\mathcal{E}_0}$ for the Drude model with parameters $a = 10$ and $100$ nm, as well as constant conductivity with $\eta = \eta_\gr$. The left panel exhibits the region of small velocity $v\ll 1$, while the right panel shows cases the entire range of $v$. The solid lines correspond to two planes with equal conductivity, and the dashed lines represent a scenario where one plane is stationary with perfect conductivity. It is worth noting that as the inter-plane distance increases, the curves approach the scenario of constant conductivity.}
	\label{fig:edr}
\end{figure}
The Casimir energy is numerically evaluated and presented in Fig.\,\ref{fig:edr}. It can be observed that as the inter-plane distance increases, the energy approaches that of the constant conductivity case (represented by the blue lines), as expected.

To compare our results with the analysis of graphene in Ref.\,\cite{Antezza:2023:Cfmg}, let us consider the case of two graphene sheets. The energy initially increases with velocity and then becomes negative, reaching a maximum at $v_c = v_F + (ma)/2$. When $m = v_F =0$, the region with positive energy disappears. Conversely, for two planes with constant conductivity, the energy is always negative for all values of velocity. Both cases exhibit a velocity correction $\sim v^2$. Regarding the inter-plane distance, different behaviors are observed. For two planes with constant conductivity, the energy has a dependence of $\sim 1/a^3$ for any distance, while for graphene, this dependence is only evident for large distances. This is expected, as the constant conductivity model for graphene is valid for large inter-plane distances. In the case of a system consisting of a perfect conductor and graphene, the energy is zero up to a specific velocity, while in the aforementioned scenario, a quadratic behavior is observed at the beginning. The same conclusion holds for the Drude model of conductivity \eqref{eq:Drude}. For large inter-plane distances, both models closely align, while for small distances, a weak dependence on distance is observed.

\section{Conclusion} \label{sec:3}

In this paper, we investigated the normal (perpendicular to the planes) Casimir force between two conductive planes with an isotropic conductivity that are moving laterally with a relative velocity $v$. Within the framework of scattering theory, the main challenge lies in determining the conductivity of a moving plane in a laboratory frame. In a co-moving frame, the isotropic conductivity is represented by a coefficient in Ohm's law, $\bm{J}' = \sigma' \bm{E}'$, where $\bm{E}'$ and $\bm{J}'$ denote fluctuations in the electric field and corresponding current density. Transforming this relation to the laboratory frame, where the first plane is at rest, is not a trivial task. A simple approach \cite{Starke:2015:Faem,Starke:2016:RcOl}  to address this issue is to start from the linear relation between current density and electromagnetic vector potential, $J^\mu = \Pi^\mu_\nu A^\nu$, which is commonly used in plasma physics \cite{Melrose:2008:Qpup}. Here, $\Pi^\mu_\nu$ represents the linear response tensor. A similar approach was employed in Ref.\,\cite{Bordag:2009:CipcgdDm}, where the polarization tensor served as the linear response tensor $\Pi^\mu_\nu$. 

Even in the case of constant conductivity, the transformation of conductivity does not have a simple form \cite{Starke:2016:RcOl}. A similar methodology was applied in Ref.\,\cite{Antezza:2023:Cfmg} where the linear relation for current and electromagnetic potential, as well as the conservation of boundary conditions, were utilized. To obtain isotropic conductivity, we adapted the expressions derived for graphene, taking the limits $v_F\to 0$ for Fermi velocity and $m \to 0$ for mass gap.  With these limits, the conductivity tensor in the co-moving frame becomes diagonal. In the laboratory frame, it has the form \eqref{eq:eta2}. Using these tensors, the Casimir energy can be calculated using expression \eqref{eq:XY}, originally derived in Ref.\,\cite{Fialkovsky:2018:Qcrcs}. 

The expressions obtained for two conductive planes \eqref{eq:E12}, and for the system (perfect conductivity)/(constant conductivity) involve two small dimensionless parameters: the velocity of the plane $v = v/c$ and the conductivity $\eta = 2\pi \sigma$ (dimensionless for 2D systems). In the case where $v\ll \eta \ll1$ the relative energy correction due to velocity is approximately given by $\sim (v/\eta)^2$ \eqref{eq:v2}. This quadratic dependence is common for normal force and different directions of motion \cite{Bordag:1984:Ccesfsnbc,Bordag:1986:CEUMM,Maslovski:2011:Crmm,Antezza:2023:Cfmg}.  The energy dependence on the inter-plane distance is $1/a^3$ for any distance, which is typical for the constant conductivity case \cite{Khusnutdinov:2014:Cescc}, as the constant conductivity model is valid for large distances where the Casimir regime is satisfied. The Drude model of conductivity shows similar behavior of the system with a weak dependence on the inter-plane distances (see Fig.\,\ref{fig:edr}). 

The constant conductivity model, discussed in Refs.\,\cite{Khusnutdinov:2014:Cescc,Khusnutdinov:2015:Cescp,Khusnutdinov:2016:Cescp,Khusnutdinov:2019:cei2dmss}, accurately describes the Casimir effect for graphenes. However, in the case of the normal force considered in this paper, there is a significant qualitative difference. When the mass gap $m$ is non-zero $m\not = 0$, the velocity correction is positive up to its maximum value at $v= v_F + am/2$, whereas the constant conductivity model gives a negative correction. As stated in Ref.\,\cite{Antezza:2023:Cfmg}, spatial dispersion plays a crucial role in the Casimir effect. For low conductivity values, the Casimir effect exhibits a linear dependence on conductivity in the constant conductivity case, whereas it becomes quadratic when spatial dispersion of conductivity is taken into account.

The next planned investigation concerns the quantum friction in the case of constant and isotropic conductivity. It is anticipated that the magnitude of this force is significantly smaller by orders of magnitude \cite{Volokitin:2007:Nrhtnf} compared to the normal force. Nevertheless, this question is of importance from an academic standpoint due to the wide range of results obtained for quantum friction.

\section*{Acknowledgments}
NK was supported in part by the grants 2022/08771-5, 2021/10128-0 of S\~ao Paulo Research Foundation (FAPESP).
%

\end{document}